\documentclass[11pt,twocolumn]{article}

\usepackage[a4paper,margin=1in]{geometry}
\usepackage{amsmath,amssymb}
\usepackage{graphicx}
\usepackage[hidelinks]{hyperref}
\usepackage{todonotes}
\usepackage[svgnames]{xcolor}
\usepackage{enumitem}
\usepackage{cuted}   
\usepackage{multicol}

\usepackage{tcolorbox}
\tcbuselibrary{breakable}

\usepackage{authblk}
\newtcolorbox{perspectivebox}[2][]{
  breakable,
  colback=gray!5,
  colframe=black!70,
  title=\textbf{#2},
  fonttitle=\bfseries,
  left=6pt,right=6pt,top=6pt,bottom=6pt,
  #1
}

\newenvironment{perspectivebox*}[1]{%
  \begin{strip}
  \begin{perspectivebox}{#1}
}{%
  \end{perspectivebox}
  \end{strip}
}

\usepackage[
  backend=biber,
  style=numeric-comp, 
  natbib=true,
  maxcitenames=2,
  maxbibnames=99,
  sorting=none,
]{biblatex}

\title{Functional Whole-Brain Models: A New Framework for Unifying Brain Structure and Cognitive Function}

\author[1,2]{Mario Senden}
\author[3]{Leonardo Dalla Porta}
\author[4]{Jan Fousek}
\author[5]{Jorge F. Mejias}
\author[6]{Gorka Zamora-L\'opez}

\affil[1]{Department of Cognitive Neuroscience, Faculty of Psychology and Neuroscience, Maastricht University, Maastricht, the Netherlands}

\affil[2]{Maastricht Brain Imaging Center, Faculty of Psychology and Neuroscience, Maastricht University, Maastricht, the Netherlands}

\affil[3]{Institute of Biomedical Research August Pi i Sunyer (IDIBAPS), Barcelona, Spain.}

\affil[4]{Central European Institute of Technology (CEITEC), Masaryk University, Czech Republic.}

\affil[5]{Cognitive and Systems Neuroscience Group, Swammerdam Institute for Life Sciences, University of Amsterdam, Amsterdam, the Netherlands.}

\affil[6]{Institute of Computer Science, The Czech Academy of Sciences, Prague, Czech Republic.}

\begin{document}
\maketitle

\begin{strip}
\begin{abstract}
Contemporary computational neuroscience features two prominent modeling traditions. Bottom-up whole-brain modeling (WBM) builds biophysically detailed simulations of brain structure and dynamics, whereas top-down neuroconnectionism optimizes deep neural networks for functional performance. Each has achieved remarkable success yet remains incomplete with WBMs lacking functional competence and neuroconnectionist models showing limited biological grounding. Here we propose functional whole-brain models (fWBMs) as a unified modeling paradigm that integrates structural and dynamical realism with task-performing capacity. fWBMs are defined by four minimal criteria: structural grounding in empirical connectomes and regional biology, continuous-time dynamical realism, functional competence across cognitive domains, and mappable observables to neuroimaging, electrophysiologcal and behavioral data. To formalize this integration, we establish a three-pillar roadmap across short-, mid-, and long-term horizons, and outline the scientific and clinical opportunities this paradigm enables. We argue that the disciplined pursuit of this integrative vision will generate the tools, common language, and cross-scale hypotheses needed to advance our understanding of the brain.
\end{abstract}
\end{strip}

\clearpage

\section*{A Field Divided}

A central challenge of computational neuroscience is to explain how the brain's biological machinery gives rise to cognition and behavior~\cite{Piccinini2014Foundations, Chirimuuta2013MinimalModels}, but the immense complexity of the brain has led the field to focus on largely isolated brain structures and functions~\cite{Machens2012Building, Senden2026EvolvingLandscape}. This has resulted in a split into two largely independent traditions~\cite{Piccinini2014Foundations} where one, a bottom-up approach, pursues mechanistic fidelity by building detailed simulations from biological components~\cite{Eliasmith2014LargeScaleModels, Djurfeldt2008LargeScaleModeling}, and a top-down approach that seeks to capture functional capacities with neurocomputational algorithms~\cite{Senden2023ModularIntegrative}. Both have achieved remarkable success, yet each remains incomplete~\cite{Senden2023ModularIntegrative}. For instance, a biophysically detailed, multi-scale, model of the entire cortex is able to reproduce neural activation profiles observed with several experimental modalities and across scales without enabling the system to perform any meaningful behavior~\cite{Pronold2024MultiscaleModel,nakagawa2014delays}. Conversely, a unified acoustic-to-speech-to-language embedding model can accurately predict neural activity across the language processing hierarchy in everyday conversations, yet its underlying computational architecture does not reflect biology~\cite{Goldstein2025UnifiedAcoustic}. As a result, we possess models of the brain's machinery and models of the mind's abilities, but no principled bridge between them. This limits our fundamental understanding of the brain and our ability to treat its disorders.

Fortunately, computational neuroscience now stands at a critical inflection point where this fragmentation can be overcome. A convergence of data and methods increasingly provides the means to build models that are both biologically realistic and functionally performant. An unprecedented wealth of available data from cellular atlases~\cite{Rood2025HumanCellAtlas} to whole-brain structural connectivity~\cite{VanEssen2012Human} and ecological tasks~\cite{Grauman2022Ego4D} provides empirical grounding, while neuroinformatics offers the tools to integrate these multi-scale datasets~\cite{Luo2024multimodal, Bjaalie2008understanding}. Simultaneously, advanced parameter optimization methods enable training biologically detailed models on ecologically valid tasks, with gradients flowing through ordinary and stochastic differential equations~\cite{Chen2018NeuralODE, Liu2019NeuralSDE, neftci2019surrogate} or discrete spiking events~\cite{Bellec2020Solution, Wunderlich2021EventBased}. These developments allow us to move beyond the trade-off between realism and relevance. Here, we take a conceptual step toward unifying bottom-up and top-down modeling approaches to form functional whole-brain models.
\section*{Two Roads to the Brain}

Within the bottom-up framework, whole-brain modeling emerged as a prominent research paradigm in the late 2000s, alongside the development of macro-connectomics \cite{Hagmann2008Structural,Rubinov2010Complex,Griffiths2021WholeBrain}. Typically, this paradigm begins by parcellating the brain into regions, extracting large-scale structural connectivity using techniques such as diffusion magnetic resonance imaging, and then coupling nodes that capture the local dynamics of brain regions according to these connections~\cite{Breakspear2017DynamicModels,Griffiths2021WholeBrain}. By simulating the resulting network we reproduce signals of brain activity, similar to those empirically recorded, that help investigate how complex spatiotemporal brain dynamics arise from first principles~\cite{Zhou2006Hierarchical,Griffiths2021WholeBrain,Vohryzek2022BrainStates}. This allows us to study how brain signals are altered in different natural states (e.g., wakefulness or sleep)~\cite{Perez2020Sleep,goldman2023simulation}, or due to brain damage~\cite{Lopez2021Stability,gaglioti2026slow}, disease~\cite{Dallmer2023Epilepsy,Patow2023Alzheimers} or under pharmacological modulation~\cite{Vohryzek2022BrainStates,sacha2025computational}. In particular, whole-brain models (WBMs) offer an excellent principled framework to investigate how the physical form of the brain shapes its intrinsic activity~\cite{Honey2009Predicting,Gomez2010Synchronization,Griffiths2021WholeBrain,dalla2025cholinergic,fousek2024symmetry}. Despite their structural and dynamical sophistication, most current WBMs are functionally limited; they primarily simulate intrinsic or perturbed brain activity but do not perform cognitive or behavioral functions~\cite{KriegeskorteDouglas2018}. Consequently, while WBMs can reveal how the brain's structural architecture and large-scale dynamics constrain the repertoire of possible brain states~\cite{Deco2012Anatomy,Senden2014RichClub,goldman2023simulation}, they typically do not yet provide an account of how specific cognitive or behavioral functions are realized.

Within the top-down framework, neuroconnectionism emerged in the mid-2010s, propelled by the success of deep learning in emulating human functional capacities~\cite{Yamins2016GoalDriven, Doerig2023Neuroconnectionist}. Neuroconnectionism proceeds by designing and optimizing deep neural network (DNN) architectures that realize functional capacities such as vision or language~\cite{Yamins2016GoalDriven, Doerig2023Neuroconnectionist}. The resulting high-performing model then becomes an object of scientific study allowing us to discover neurocomputational principles that can transform sensations into meaningful behavior~\cite{Doerig2023Neuroconnectionist}. However, the approach remains limited by its lack of biological realism~\cite{Bowers2022DeepProblemsNNVision}. Because the architectures and processing units used in most DNNs are only weakly constrained by known neuroanatomy and neurophysiology, it is difficult to determine whether a model's success reflects brain-like principles or artifacts of its engineering constraints~\cite{Richards2019Deep}.

These two paradigms clearly offer complementary insights, with WBM's emphasis on biological structure and intrinsic dynamics and neuroconnectionism's emphasis on functional competence and representational transformations. Their integration is therefore a promising and increasingly necessary direction for computational neuroscience.
\section*{The First Bridges}

Unsurprisingly, a growing body of research has begun to move this integration forward. For example, Damicelli et al.~\cite{Damicelli2022Brain} recently explored whether empirically measured whole-brain connectomes could serve as architectural blueprints for echo state networks. By replacing the random recurrent connectivity typically employed in reservoir computing with fixed connections derived from macaque, marmoset, and human brain data, they evaluated performance on various memory benchmark tasks. Their work demonstrates that the connectome can serve as a structural prior, underscoring the feasibility of mapping neural architecture to functional networks and providing a baseline for connectome-constrained computation using static biological topologies.

Expanding on the use of structural priors, Goulas et al.~\cite{Goulas2021Bio} moved to trainable architectures by integrating whole-brain anatomical constraints into recurrent neural networks. Unlike the reservoir approaches of Damicelli et al., their models utilized backpropagation-through-time, allowing bio-instantiated connections to serve as an inductive bias that guides the learning of working memory tasks. Their systematic evaluation provided insights into the interplay between architecture and connectivity, finding that while biological topology significantly shapes the learning process, the emergence of functional capacity remains dependent on the subsequent diversification of connection weights. By demonstrating that structural brain data can effectively constrain a learning network, this work establishes a viable path toward brain-inspired design while simultaneously highlighting the value of task-driven optimization to achieve functional competence.

In parallel, recent work by Mejias et al.~\cite{Mejias2022Mechanisms} has shown how intrinsic dynamics, shaped by detailed anatomical features, can give rise to emergent function. Their large-scale model of the macaque cortex utilized 30 interconnected areas, each represented by mean-field models of excitatory and inhibitory populations constrained by tract-tracing connectomes, hierarchical excitability gradients, and laminar projection patterns. The model mechanistically explained how innate working memory capacity could be maintained through distributed, self-sustained activity across multiple cortical areas without the need for explicit task-specific optimization. By showing that complex cognitive phenomena can arise from the interplay of biological structure and dynamics alone, this study underscores the potential for whole-brain models to reveal fundamental principles of brain function.

The computational utility of intrinsic dynamics was further explored by Effenberger et al.~\cite{Effenberger2025Functional} through topologically abstract oscillatory recurrent networks. By constructing networks in which each unit was a damped harmonic oscillator, they demonstrated that oscillatory dynamics may provide a powerful resource for functional performance. Trained on pattern recognition tasks, these models consistently outperformed conventional recurrent networks in learning speed, noise robustness, and parameter efficiency. Their identification of wave-based computation as a core principle offers a compelling functional justification for the ubiquitous oscillatory activity in the brain and proves that rich intrinsic dynamics are fully compatible with, and even beneficial for, gradient-based learning.

The integration of biological principles at a system-wide scale has also been common practice within the neuroconnectionist tradition. For instance, Kubilius et al.~\cite{Kubilius2018CORnet} exemplified a successful integration of biological architecture with task-driven optimization within the primate ventral visual stream. Their CORnet family of deep neural networks incorporated a shallow, four-area hierarchy mirroring the biological visual system (V1, V2, V4, and inferior temporal cortex) along with intra-area recurrence. Optimized for core object recognition, these models achieved high predictive power against neural and behavioral data while abstracting away intrinsic dynamics in favor of standard deep learning activations. This work demonstrates that system-level constraints, such as hierarchical depth and recurrence, can serve as powerful inductive biases, enabling models to achieve functional capabilities comparable to much deeper networks while remaining neurobiologically plausible.

The explanatory power of biological scaling was further explored by Weidler et al.~\cite{WeidlerLateral} in the context of the primate frontoparietal system. Their model incorporated several neuroanatomical constraints, including a connectome-based network architecture and, crucially, cell count-based capacity scaling to reflect biological limitations. When trained via deep reinforcement learning to perform complex motor manipulation, a clear functional hierarchy emerged spontaneously: the lateral prefrontal cortex specialized in abstract goal planning, while the premotor cortex handled motor sequencing. This work demonstrates that simple architectural scaling can, when grounded in quantitative biological data, serve as a powerful inductive bias for the emergence of sophisticated cognitive hierarchies in embodied systems.

Complementing these data-driven approaches, Granier et al.~\cite{granier2025multihead} provided a top-down functional rationale for specific biological motifs by showing how the canonical thalamo-cortical circuit is uniquely suited to implement the logic of multihead self-attention. By mapping the components of a key-value memory system onto the specific laminar structure and thalamic projection patterns of the brain, they established a conceptual blueprint for how high-performance algorithms can guide our understanding of biological form. This approach suggests that cortical organization is not merely a collection of constraints but a physical implementation of efficient computational motifs, offering a theoretical bridge that links algorithmic requirements to biological implementation.

A significant step toward the convergence of these principles is represented by de Leeuw et al.~\cite{DeLeeuwDeep}, who introduced a biophysics-informed deep learning approach using interconnected cortical columns. Each column was represented by a dynamic mean-field model incorporating distinct layers and populations, successfully bridging the meso-scale biophysics of traditional WBMs with the goal-driven optimization of neuroconnectionism. By demonstrating that such biologically grounded units can be robustly trained for function while reproducing realistic neural dynamics and connectivity motifs, this work provides a proof-of-concept for the fWBM vision. It proves that sophisticated dynamical units can be successfully embedded within a task-driven framework, offering a scalable path toward unifying biological realism with functional competence.

Together, these diverse efforts demonstrate that the gap between biological form and functional competence is bridgeable, yet they remain largely fragmented across different scales and methodologies. This underscores the need for a unified framework that can systematically integrate these breakthroughs into a single, coherent modeling paradigm.
\section*{Functional Whole-Brain Models}

\begin{figure*}[t!]
    \centering
    \includegraphics[width=\textwidth]{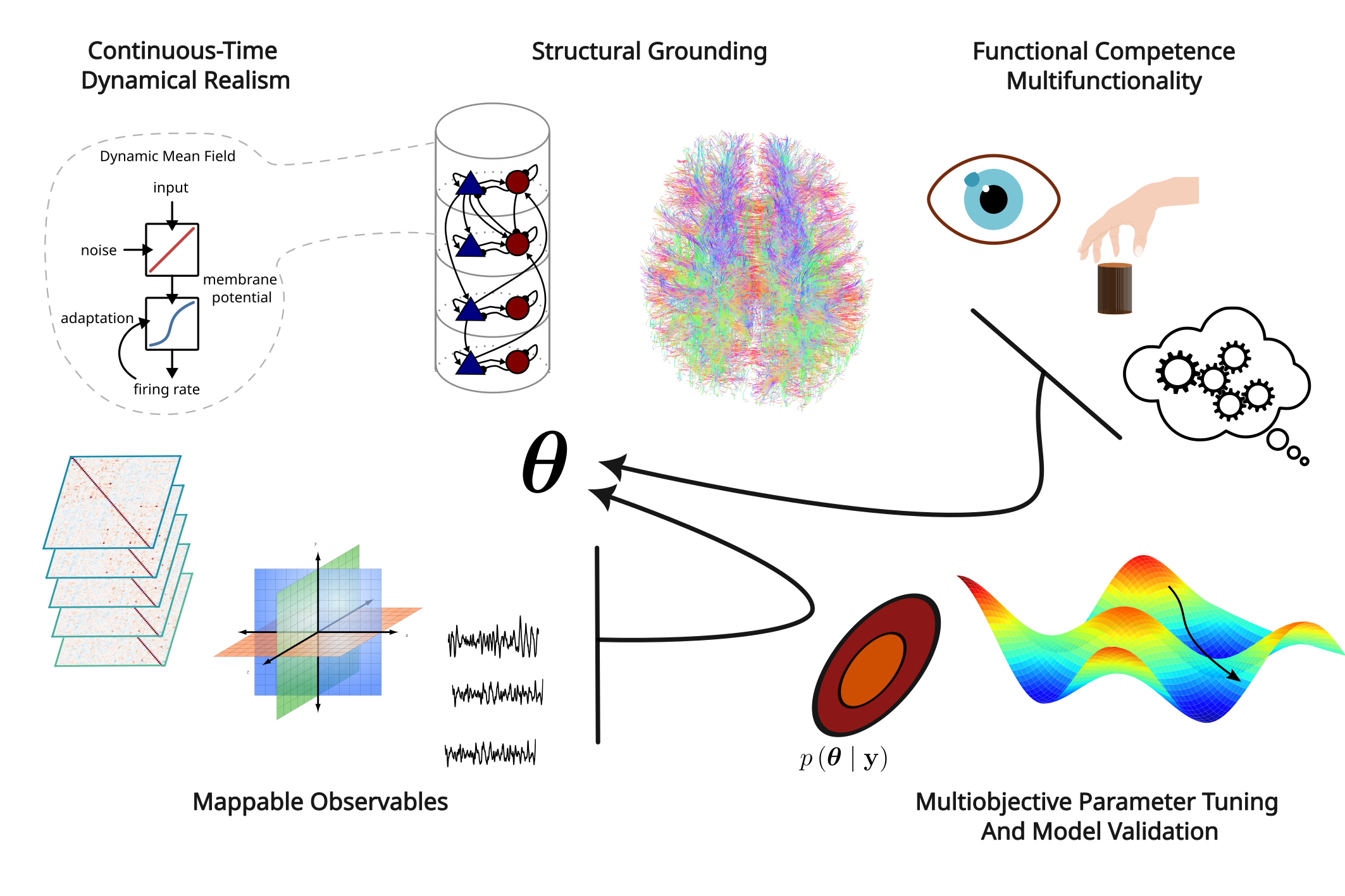}
    \caption{\textbf{The functional whole-brain modeling paradigm.} fWBMs are defined by four minimal criteria, illustrated here as an integrated system. \textit{Structural grounding} (top center): model architecture is derived from empirical meso- and macro-scale structural connectivity together with region-specific biological properties. \textit{Continuous-time dynamical realism} (top left): local dynamics are governed by differential equations with neurophysiologically interpretable state variables, exemplified by a dynamic mean-field (DMF) unit. \textit{Functional competence and multifunctionality} (top right): the model actively processes ecologically valid inputs and generates overt and covert behavior. \textit{Mappable observables} (bottom left): internal dynamics generate signals directly comparable to empirical neuroimaging modalities such as fMRI, EEG, and LFP as well as internal representations comparable to those observed in biological systems. Functional competence requirements and mappable observables can be used for model fitting and validation. Model fitting is achieved through \textit{multi-objective parameter tuning} (bottom right), in which Bayesian inference initializes parameters at biologically plausible operating points, and gradient-based optimization (regularized by biological constraints) refines model parameters toward task performance.} 
    \label{fig:fWBMdefined}
\end{figure*}

To realize this synthesis, we propose functional whole-brain models (fWBMs) as a unified modeling paradigm. At their essence, fWBMs are large-scale dynamical systems whose activity is constrained by the brain's anatomical network, yet which possess the inherent capacity to perform a plethora of perceptual, cognitive and motor functions. By unifying the structural and dynamical realism of bottom-up modeling with the task-performing power of neuroconnectionism, this paradigm provides a principled platform for understanding how the brain's physical implementation, its dynamics and its cognitive abilities are inextricably linked within a single coherent framework. Importantly, fWBMs are not intended as fully biophysical replicas at every spatiotemporal resolution, nor are they a guaranteed, unique explanation for all neural data. Instead, they represent a meaningful complementary approach designed to facilitate the testing of multi-scale hypotheses through a set of four minimal and principled criteria.

First, fWBMs require \textbf{structural grounding}. Unlike the abstract or fully connected topologies common in deep learning \cite{Bowers2022DeepProblemsNNVision}, the architectures of fWBMs must be derived from empirical structural connectivity and regional biological properties \cite{Griffiths2021WholeBrain}. This includes macro-scale connectomes defining white-matter pathways and transmission delays, as well as meso-scale metadata such as region-specific cell densities, receptor distributions and laminar motifs \cite{bazinet2023towards}. In this view, the neuroanatomical structure constitutes an inductive bias that constrains the repertoire of possible functional capacities and shapes the geometry of neural representations \cite{KriegeskorteDouglas2018}. 

Second, fWBMs must exhibit \textbf{continuous-time dynamical realism}. The system should be formulated as a set of ordinary or stochastic differential equations \cite{Breakspear2017DynamicModels}, where state variables represent neurophysiologically interpretable quantities such as population firing rates. Crucially, these models must be capable of generating self-sustaining network-level dynamics even in the absence of task input \cite{Deco2011Emerging}. By maintaining a realistic resting-state operating point, fWBMs ensure that cognitive functions emerge from a biologically plausible dynamical regime characterized by, for example, oscillations, attractors and metastable states \cite{Deco2012RestingCriticality}.

Third, the paradigm demands \textbf{functional competence and multifunctionality}. fWBMs move beyond the simulation of passive or perturbed brain states to actively processing ecologically valid stimuli and generating behavior. While early models may focus on specific functions, the ultimate goal is the creation of multifunctional models where a single neuroanatomical substrate can flexibly engage in capacities originating in different cognitive domains, such as from visual recognition to auditory processing or motor planning.

Finally, fWBMs must produce \textbf{mappable observables}. Their internal dynamics must give rise to signals that are directly comparable to empirical neuroimaging and electrophysiology, such as functional magnetic resonance imaging (fMRI), electroencephalography (EEG) and electrocorticography (ECoG). This ensures the model remains anchored to measurable neurobiology while performing functions. Validation further extends beyond neural signals to include a dual accountability to behavioral benchmarks, requiring the model to match the reaction times, error patterns and psychometric curves of biological organisms to ensure that the mechanisms generating the behavior align with biological observations. 

Realizing this vision requires an architectural synthesis based on the principle of compositional assembly. fWBMs are constructed from standardized, differentiable modules including cortical functional units, such as canonical microcircuits parameterized by regional biological data and their specialized subcortical and cerebellar counterparts. Crucially, these components are integrated as biologically grounded dynamical (sub)systems rather than generic connectionist activation functions, allowing the approach to capture the specialized roles of, for example, thalamic gating or cerebellar timing within the larger global network. This assembly biases specific functions to arise from the interplay of localized regional processing and distributed integration across the connectome, ensuring that computation remains an emergent network property rather than a property of isolated components.

The technical operationalization of fWBMs provides a bridge between structural and functional requirements. By formulating the brain as a differentiable system of continuous-time differential equations, the paradigm enables the use of gradient-based optimization. This allows us to implement a two-stage strategy where the system is first placed into a biologically plausible operating regime using informed priors and established whole-brain modeling techniques, such as Bayesian model inversion, to target intrinsic dynamical properties \cite{Hashemi2025}. Following this initialization, gradient-based optimization propagates gradients through the governing differential equations \cite{Chen2018NeuralODE}, allowing parameters to be fine-tuned for task competence while architectural and dynamical constraints can be incorporated as regularizers. This multi-objective strategy balances task performance with biological realism, closing the loop between physical form, emergent dynamics and functional capacities to provide a unified platform for multi-scale hypothesis testing.

While the envisioned fWBMs are primarily motivated by human neuroscience, they are inherently species-agnostic. They can serve as a principled paradigm applicable to any organism for which structural connectivity and functional data exist, such as rodents and non-human primates. This broad scope allows for cross-species comparisons and the validation of human-like dynamics against invasive recordings in animal models.
\section*{The Prize of Unification}

fWBM could provide a powerful platform for addressing a new set of scientific and clinical questions. By combining structural and dynamical realism with task performance, fWBMs may enable the exploration of how the brain's physical implementation and its cognitive abilities are inextricably linked. This shift could expand large-scale modeling from a focus on reproducing intrinsic activity patterns into a teleological mechanistic paradigm for understanding the mind.

\subsection*{A testbed for theories and hypotheses}

The fWBM paradigm would, for instance, enable us to investigate how the brain's physical network topology shapes the dynamics of its representational geometries~\cite{KriegeskorteDouglas2018} and serve as testbeds for various hypotheses. For example, we could analyze how specific network properties of the connectome affect the encoding and transformation of information during the execution of tasks over ecologically valid stimuli. Similarly, it would allow us to probe the computational purpose of brain rhythms~\cite{buzsaki2023brain,Effenberger2025Functional} and evaluate whether oscillatory dynamics support information routing~\cite{Palmigiano2017Routing} in temporally extended tasks under large-scale constraints, such as conduction delays. The paradigm further may enable the study of higher-level organizing principles. For example, fWBMs can be used to compare candidate mechanisms of flexible cognitive control, a goal that has so far remained out of reach for either paradigm alone~\cite{VanHolk2024CogFlex}, such as testing how modulation of putative control hubs or subcortical gating circuits drives task-dependent state transitions. This could provide a path to understanding the mechanisms of cognitive flexibility and goal-directed behavior. Finally, it might allow for exploring the functional implications of metabolic efficiency. By optimizing a model for both task performance and minimal energy cost, we may discover which computational strategies arise as a consequence of the need to be efficient~\cite{Grytskyy2021Learning,Deneve2016EfficientCodes}. Ultimately, this synthesis could demonstrate how localized circuit motifs and distributed integration across the connectome jointly give rise to emergent computation.

\begin{figure*}[t]
\centering
\include{boxes/box_01}
\end{figure*}

fWBM further offers a promising scenario for testing grand theories in neuroscience, such as the Bayesian brain hypothesis, predictive coding and the critical brain hypothesis. Both predictive coding and the Bayesian brain hypothesis posit that the brain constantly generates and updates a mental model of the environment to predict incoming sensory input~\cite{Rao1999Predictive,Friston2010FreeEnergy,Salvatori2023BrainInspired}. The brain is assumed to minimize prediction error over sensory inputs. In this view, perception is largely driven by top-down predictions derived from internal beliefs, which in turn are refined by prediction errors~\cite{Sprevak2024Predictive, Bottemanne2025BayesianBrain, Friston2005CorticalResponses,Salvatori2023BrainInspired,Rao1999Predictive}. The Bayesian brain hypothesis couches this process in a probabilistic paradigm where the brain represents the world probabilistically and updates its representations according to Bayesian principles~\cite{Friston2005CorticalResponses, Bottemanne2025BayesianBrain,Friston2009FreeEnergy}. Although powerful, testing the mechanistic claims of predictive coding and the Bayesian brain hypothesis has been challenging. While neuroconnectionist models can implement the core algorithms, their abstraction from biology renders deeper validation difficult. Such models typically lack critical neuroanatomical details, such as the distinct cortical layers hypothesized to handle prediction and error signals~\cite{bastos2012canonical}. Furthermore, they struggle to capture the continuous, recurrent interactions between top-down predictions and bottom-up errors~\cite{Hosseini2020HierarchicalPredictiveCoding, Salvatori2023BrainInspired}. Similarly, neuroconnectionist models lack neuromodulatory-like effects (e.g., time-varying acetylcholine), which constitutes a significant gap given that the Bayesian brain hypothesis postulates neuromodulator-driven mechanisms for dynamically updating prior beliefs and context-dependent modulation of prediction error precision~\cite{Williams2021Epistemic, Lange2023BayesianCoding}. fWBMs can incorporate these biological features, allowing us to test how the brain might flexibly adapt its Bayesian inferential processes. While these models will not resolve all open questions surrounding these theories, they might provide a more comprehensive paradigm for testing them.

With respect to the critical brain hypothesis, similar concerns regarding testability persist, albeit in a different form. The critical brain hypothesis proposes that the brain operates near a critical point of a second-order phase transition~\cite{Chialvo2010Emergent, Hesse2014Criticality,Beggs2022CriticalCortex}. Since systems that operate near such a point have been shown to exhibit favorable computational properties such as efficient information transmission, the critical brain hypothesis postulates that brain dynamics self-organize toward criticality to exploit these computational advantages~\cite{OByrne2022Critical, Hengen2024Criticality, Hesse2014Criticality,Beggs2022CriticalCortex}. 
While intriguing, testing functional claims of the critical brain hypothesis has been difficult. There is a form-function gap that creates a crucial distinction between the two versions of the hypothesis. The phenomenological critical brain hypothesis, which focuses on explaining the scale-free statistics of the brain, can be effectively probed with the whole-brain modeling approach; these models can reproduce the intrinsic dynamics of the brain and show that they have properties consistent with near-critical dynamics~\cite{Deco2012RestingCriticality,haimovici2013brain,rabuffo2025connectome,myrov2026hierarchical}. However, this approach fails to test the stronger claim that the brain operates at criticality because it is behaviorally beneficial~\cite{cambrainha2025criticality,muller2025critical}. Because current whole-brain models do not perform tasks or interact with an environment, they cannot establish whether criticality genuinely improves cognitive performance. Measures of information capacity remain proxies without a ground truth for what is useful for an organism. By embedding a whole-brain model within a task-driven neuroconnectionist paradigm, we may be able to test the functional critical brain hypothesis more directly than before. It would allow us to assess whether optimizing a model for a cognitive task naturally tunes its parameters toward a critical state~\cite[e.g., ][]{safavi2024signatures} and whether behavioral performance is indeed enhanced when the model operates near or at a critical point, as suggested by human studies~\cite{muller2025critical}.

\subsection*{Real-life implications of functional whole-brain models}

Lastly, fWBMs hold the potential to significantly advance clinical and translational neuroscience by providing a causal bridge between low-level pathophysiology and high-level functional deficits. This could open the door to powerful new paradigms for understanding and treating brain disorders. For example, it may allow for modeling abnormalities such as the degradation of white matter tracts in multiple sclerosis, receptor imbalances characteristic of Parkinson's disease, or the effects of amyloid-beta plaques on synaptic function in Alzheimer's~\cite{DecoKringelbach2014Great}. By tasking models characterized by specific pathologies with cognitive tests, we might be able to build a causal chain from a precise biophysical fault to a measurable behavioral symptom. 

fWBMs could also enable new approaches in the design of neuromodulation therapies. Currently, the optimization of stimulation parameters aims to restore a predefined healthy brain state based on measured activity patterns~\cite{perl2023model}. An fWBM might allow for a function-oriented approach. Instead of tuning stimulation to merely match that of a control group, it could be optimized directly to improve performance on a relevant cognitive or motor task within the model. In the long term, this approach may pave the way for digital twins that supplement a patient's structural connectome and functional measurements with their individual behavioral data~\cite{wang2024virtual}. Such a digital twin could be used to simulate disease progression, make predictions about treatment responses and test novel interventions virtually before administering them to the patient~\cite{Hashemi2025}.

\section*{A Roadmap to Functional Whole-Brain Models}

The vision for fWBMs is admittedly ambitious, but not out of reach. To make it concrete and feasible, we propose a roadmap built on three parallel, mutually-informing research pillars. We propose to articulate the goals for each pillar across a realistic timeline, distinguishing between \textbf{short-term}, \textbf{mid-term} and \textbf{long-term} objectives. This staged approach ensures that valuable scientific insights are generated at every step, long before the ultimate vision is realized. In the short-term, we envision system-level models (e.g., the ventral visual stream) that can be trained on a limited set of tasks. In the mid-term, we expect integrated, multi-system models (e.g., visual and auditory cortical streams) capable of more complex cognitive tasks and the ability to flexibly switch between them. The highly aspirational long-term goal is the realization of general multifunctionality within a single, unified whole-brain architecture, capable of solving a diverse suite of cognitive tasks and serving as a virtual testbed for exploring how global dynamics support flexible behavior.

The first pillar is dedicated to creating a comprehensive, open-source library of standardized functional units and automated scaffold generation tools for constructing fWBMs. In the \textbf{short-term}, the focus should be on developing cortical functional units; i.e., differentiable modules of canonical microcircuits and automated tools that translate atlas and connectivity data into model graphs. In the \textbf{mid-term}, these foundational elements should be refined with greater biological detail, including region-specific interlaminar connectivity patterns. The \textbf{long-term} goal is a community-wide effort to develop and validate standardized functional units for subcortical structures and the cerebellum, extending the library beyond the cortex. Short-term functional units and anatomical scaffolds will already allow researchers to test how local circuit motifs and structural constraints shape representational geometry, dynamic signatures and attractor regimes during simple tasks.

The second pillar concerns the specification of objectives and optimization algorithms that endow whole-brain models with task-relevant computational capacities. Its central goal is to ensure that optimization procedures yield brain-like yet functionally meaningful dynamics. \textbf{Short-term} goals should focus on tractable first steps, such as using supervised learning with adjoint methods and simple, linear readout probes on specific regions of interest to validate model representations (e.g., against RDMs or simple classification) in response to sensory inputs. Importantly, models must be tuned through a two-stage strategy where the system is first placed into a biologically plausible operating point using informed priors to target intrinsic dynamical properties, followed by gradient-based optimization that fine-tunes parameters for task performance while maintaining biological realism. We refer to this as multi-objective functional connectome tuning. In the \textbf{mid-term}, the field should graduate to more powerful unsupervised and self-supervised objectives and train models on more complex cognitive tasks that require flexible control and multi-tasking. The \textbf{long-term} ambition is to use a combination of self-supervised and reinforcement learning signals to train whole-brain models to solve a restricted set of ecologically relevant tasks. It is important to note that these gradient-based optimization approaches cannot be considered models of biological learning, but merely optimization tools. Integrating more biologically plausible learning rules thus constitutes another key long-term challenge for this pillar. 

The third pillar addresses the architectural and systems-level challenge of assembling fWBMs from the components described in the first pillar and the functional optimization strategies developed in the second. Its focus is on how heterogeneous modules trained under different objectives can be integrated into a single coherent system. The \textbf{short-term} proof-of-concept stage will rely on methods like co-simulation and partial training, focusing learning on a small, tractable subset of the model to create the first testable hybrid models. The \textbf{mid-term} stage involves engineering larger-scale modular systems in which only selected modules are trainable, building on pre-trained and fixed modules as needed. This progression culminates in the \textbf{long-term} stage, which focuses on constructing a multifunctional whole-brain model capable of solving a diverse suite of perceptual and cognitive benchmarks within a unified anatomical substrate. This goal must be approached realistically; we do not envision training monolithic models from scratch, but rather assembling them from a community-governed registry of pretrained and fixed modules with light-touch finetuning to achieve general functional competence.

We believe this roadmap offers immediate and tractable entry points for researchers across disciplines. No single community needs to build an entire fWBM; instead, each can contribute components that fit naturally within its expertise. For \textbf{computational neuroscientists}, the entry points are developing novel cortical functional units with varying levels of biological detail, implementing more efficient and stable solvers for stiff or delayed differential equations, or contributing to the open-source platforms that will underpin this entire effort. For \textbf{cognitive and systems neuroscientists}, the opportunities lie in designing new behavioral tasks that can probe the limits of these models, developing standardized forward models to translate internal dynamics into observable neural signals and providing rich experimental datasets (fMRI, EEG, ECoG) to constrain and validate fWBMs. For \textbf{data scientists and machine learning engineers}, the tasks range from developing novel parameter optimization and self-supervised learning techniques suited to high-dimensional, sparsely constrained dynamical systems to exploring new methods for multimodal data fusion and model regularization. Collaboration across these disciplines will be the driving force that propels the roadmap forward.

\begin{figure*}[t!]
\centering
\input{boxes/box_02.tex}
\end{figure*}

Across all stages, biological realism must be balanced against numerical tractability. Early fWBMs will necessarily employ simplified, coarse-grained dynamics that retain essential organizational principles while avoiding the computational burden of full biophysical detail. We will encounter several technical challenges along the way. These include the high computational cost and numerical instability of simulating large, stiff dynamical systems; the intrinsic difficulty of credit assignment in deep, recurrent networks; the severe under-identification of models due to sparse and indirect data; and the current lack of a unified software toolchain. Because macaque and human neural data are sparse, indirect and heterogeneous, fWBMs must be interpreted as tools for constraining mechanism classes rather than recovering a single ground-truth model. These constraints motivate a pragmatic approach that emphasizes modular, partially trainable models, pre-training and fine-tuning and multi-objective functional fits. The scientific product will not be a single ground-truth model, but rather an ensemble of models.

\section*{Conclusion}

Computational neuroscience has long been split between two grand traditions, with one modeling the brain's neurobiological form and the other emulating its functional capacities. We argue that synthesizing the two paradigms of whole-brain modeling and neuroconnectionism into fWBM is a crucial next step for the field. It presents a concrete path to move toward a more mechanistic understanding of how biological structure and dynamics can give rise to cognition.

The successful development of fWBMs, following the roadmap we have outlined, has the potential to significantly advance our understanding of the brain. Grand organizing principles, from predictive coding to the critical brain hypothesis, can be instantiated and tested not as abstract algorithms, but as emergent properties of a physically grounded system. In clinical neuroscience, this approach could open new avenues for understanding disease and designing interventions. In the long term, populations of disease twins could help generate hypotheses about typical patterns of a disorder, and heavily regularized patient-specific models may one day serve as interactive priors for clinical reasoning.

Realizing this vision requires a collective reorientation. It demands that we build shared, open-source tools, create common benchmarks, and foster a culture of deep interdisciplinary collaboration. The path forward is no longer defined by a choice between biological realism and functional relevance, but by the challenge of their integration.

We do not expect to build a fully functional, autonomous whole-brain model in the next decade. The very idea may be a fantasy. However, we argue that the disciplined pursuit of this goal constitutes a fruitful way to advance computational neuroscience. By committing to this integrative vision, we will build the tools, develop a common language, and generate testable, cross-scale hypotheses necessary to move beyond our current, fragmented understanding of the brain. The prize is not the final model; it is the science we will discover along the way.
\section*{Acknowledgments}

MS was supported by the European Union’s Human Brain Project SGA3 (No. 945539).
 LDP acknowledges support from INFRASLOW (PID2023-152918OB-I00; MICIU/AEI/FEDER, UE), the European Research Council project NEMESIS (No. 101071900), AGAUR (2021-SGR-01165), and the European Union’s Human Brain Project SGA3 (No. 945539), and from the Brazilian agency CNPq (Grant No. 444500/2024-3).
 JFM was supported by funding from the European Union’s Horizon Europe Program under the specific grant agreement 101137289 (Virtual Brain Twin project), NWO NWA-ORC grant NWA.1292.19.298, and the NWO LRSI project 184.037.014 (EBRAINS-Neurotech).
 G.Z.L. was supported by ERDF-Project Brain dynamics, No. CZ.02.01.01/00/22\_008/0004643.
 JF was supported by the project \textit{A lifetime with Language: the nature and ontogeny of linguistic communication} (project ID CZ.02.01.01/00/23\_025/0008726), co-funded by the European Union.

\clearpage 
\onecolumn
\begin{perspectivebox}{Glossary}
\begin{description}[leftmargin=!,labelwidth=3.2cm]

\item[Adjoint method] A mathematical technique used to efficiently compute gradients in systems described by differential equations. In deep learning, it enables the training of models such as neural ODEs by solving a second-order differential equation backward in time, avoiding the memory overhead of storing intermediate steps and allowing continuous-time backpropagation.

\item[Anatomical Scaffold] The structural backbone of a whole-brain model, typically derived from empirical connectome data, defining the connections and distances between brain regions.

\item[Attractor] A state or set of states toward which a dynamical system tends to evolve and at which it settles if unperturbed. Common types include fixed points, limit cycles (periodic oscillations), and strange attractors (chaotic trajectories). In brain models, distinct attractors correspond to different activity patterns or cognitive states. Closely related are \textit{metastable states}: quasi-stable configurations in which the system dwells for extended periods before spontaneously transitioning to another, without permanently settling at any single attractor. Metastability is thought to underlie the flexible, context-dependent nature of cognition by enabling rapid switching among functional states.

\item[Backpropagation-Through-Time (BPTT)] The standard algorithm for training recurrent neural networks by unfolding the network's computations over time and applying the backpropagation algorithm to the resulting computational graph.

\item[Bayesian Brain Hypothesis (BBH)] A theory proposing that the brain represents information probabilistically and performs computations that approximate Bayesian inference, constantly updating its beliefs about the world based on sensory evidence.

\item[Bayesian Model Inversion] The process of inferring the parameters of a generative model from observed data by computing or approximating the posterior distribution over parameters. In whole-brain modeling, it uses methods such as variational inference or Monte Carlo sampling to identify parameter values that best explain empirical neural recordings (e.g., fMRI or EEG) given biologically motivated priors. In the fWBM context, it serves as the first stage of multi-objective functional connectome tuning, placing the model into a biologically plausible operating regime before gradient-based task optimization.

\item[Biophysics-Informed Deep Learning] A deep learning approach that builds neural networks from biologically realistic cortical microcircuits rather than abstract artificial units. Models are constructed from laminar-resolved cortical columns with empirically derived connectivity and population dynamics, capturing key neuroanatomical features such as excitatory/inhibitory populations and layer-specific interactions. These networks remain trainable using tools like the adjoint method for continuous-time optimization, enabling models that are both biologically grounded and functionally expressive.

\item[Bottom-Up Modeling] An approach in computational neuroscience that aims to build models by assembling and simulating detailed biological components (e.g., neurons, microcircuits) to see what larger-scale phenomena, such as whole-brain dynamics, emerge from their interactions.

\item[Compositional Assembly] A design principle for fWBMs where models are constructed from standardized, differentiable modules (such as CFUs) and specialized subcortical counterparts, integrated as biologically grounded dynamical systems rather than generic connectionist nodes.

\item[Connectome] A comprehensive map of neural connections in a brain. In the context of this paper, it typically refers to macro-scale white-matter pathways between brain regions, often measured using diffusion MRI.

\item[Continuous-time Dynamical Realism] A core criterion for fWBMs requiring that the system be formulated as a set of differential equations where state variables represent neurophysiologically interpretable quantities, capable of generating self-sustaining intrinsic dynamics even in the absence of task input.

\item[Cortical Functional Unit (CFU)] A proposed term for a standardized, differentiable module of a canonical cortical microcircuit, serving as a reusable building block for functional whole-brain models.

\item[Critical Brain Hypothesis (CBH)] A theory suggesting that the brain operates near a critical point of a second-order phase transition to optimize properties such as information processing, dynamic range, and computational power.

\item[Critical point] The precise value of a control parameter at which a phase transition occurs. At this point, the system shifts from one qualitative mode of behavior to another. In second-order transitions, the critical point marks a state of heightened sensitivity, where small changes can propagate widely, and the system exhibits long-range correlations and slow dynamics. It is often associated with complex, flexible, and information-rich behavior.

\item[Deep Neural Network (DNN)] A class of machine learning models with multiple layers of processing units (artificial neurons), central to the neuroconnectionism paradigm for modeling cognitive functions.

\item[Diffusion MRI] A magnetic resonance imaging technique that infers white-matter fiber tract orientations by measuring the directional diffusion of water molecules in brain tissue, which is greatest along the long axis of myelinated axons. Tractography algorithms applied to diffusion MRI data reconstruct large-scale white-matter pathways, providing the primary empirical basis for the structural connectomes used in whole-brain modeling.

\item[Digital Twin] A virtual model of a physical system. In a clinical context, a patient-specific computational brain model is used to simulate disease progression, test personalized interventions, and predict treatment outcomes.

\item[Dynamic Mean-Field (DMF) Model] A mathematical formalism that approximates the average activity of a large population of neurons, often used as the dynamical ''node" representing the dynamics of a brain region in a whole-brain model.

\item[Echo State Network (ESN)] A type of recurrent neural network with a fixed, random recurrent part (the ``reservoir"), where only the output connection weights are trained. It is a form of Reservoir Computing.

\item[Embodied Task] A task that requires a model (an agent) to interact with a simulated or physical environment, involving a closed loop of perception, action, and sensory feedback.

\item[Free Energy Principle] A theoretical framework proposing that the brain maintains internal models and continuously updates them to minimize variational free energy. Closely linked to the Bayesian brain hypothesis and predictive coding, it casts perception, action, and learning as processes of minimizing the difference between expected and actual sensory input.

\item[Functional Competence and Multifunctionality] A core criterion for fWBMs requiring models to actively process ecologically valid stimuli and generate behavior across multiple cognitive domains within a single anatomical substrate.

\item[Functional Whole-Brain Model (fWBM)] The central concept of this paper: a model that integrates the structural and dynamical realism of a whole-brain model with the task-performing, learning capabilities of neuroconnectionism.

\item[Inductive Bias] In the context of fWBMs, the principle that neuroanatomical structure (the connectome) constitutes a set of prior constraints that shapes the repertoire of possible functional capacities and the geometry of neural representations.

\item[Mappable Observables] A core criterion for fWBMs requiring that internal dynamics produce signals directly comparable to empirical neuroimaging (e.g., fMRI, EEG) and that behavioral outputs match biological benchmarks.

\item[Mechanistic Fidelity] The degree to which a model's components, operations, and dynamics correspond to the underlying biological mechanisms they aim to represent.

\item[Multi-objective Functional Connectome Tuning] A two-stage optimization strategy where a model is first placed into a biologically plausible operating regime using informed priors, followed by gradient-based fine-tuning for task performance.

\item[Neural Mass Model] A mathematical description of the collective activity of large populations of neurons, often used as the dynamical building block for regions in whole-brain models.

\item[Neural ODEs] Short for neural ordinary differential equations, these models represent neural network computations as continuous-time dynamical systems. Instead of passing data through discrete layers, a neural ODE learns a differential equation that describes how the system's state evolves over time. This framework extends deep learning to continuous domains, enabling flexible modeling of time-varying data and gradient-based training via the adjoint method.

\item[Neuroconnectionism] A top-down research paradigm that uses performance-optimized deep neural networks to model cognitive functions and discover the neurocomputational principles that support them.

\item[Parcellation] The process of dividing the brain into a set of distinct, non-overlapping regions (parcels) based on anatomical, functional, or connectivity-based criteria.

\item[Prediction Error] The difference between expected sensory input, generated by the brain’s internal model, and actual input from the environment. Prediction errors signal mismatches that drive learning and perception in many computational neuroscience theories, including predictive coding and the free energy principle. By minimizing prediction errors, the brain refines its beliefs about the world and adapts behavior accordingly.

\item[Predictive Coding (PC)] A theory of brain function which posits that the brain is a prediction machine that constantly tries to minimize the error between its top-down predictions of sensory input and the actual bottom-up input it receives.

\item[Recurrent Neural Network (RNN)] A type of neural network with feedback connections, allowing it to process sequences of data and maintain an internal state or "memory" over time.

\item[Representational Dissimilarity Matrix (RDM)] A square matrix that represents the pairwise dissimilarity (e.g., 1 minus correlation) between the neural activity patterns elicited by a set of stimuli. It is a core tool of Representational Similarity Analysis (RSA).

\item[Representational Geometry] The study of how information is organized and encoded in neural population activity. It focuses on the geometric relationships (e.g., distances or angles) between activity patterns representing different stimuli or cognitive states. Analyzing this geometry helps reveal how the brain separates, generalizes, or transforms information across tasks and modalities.

\item[Representational Similarity Analysis (RSA)] A framework for comparing how information is encoded across different systems — such as brain regions, computational models, or behavioral data — without requiring one-to-one correspondence between their individual units. RSA characterizes each system's responses to a stimulus set via a representational dissimilarity matrix (RDM) and then compares these matrices across systems using rank correlations or related metrics. This allows researchers to assess whether two systems encode information in a structurally similar way, regardless of their implementation details.

\item[Reservoir Computing] A computational framework that exploits the rich nonlinear dynamics of a fixed, high-dimensional recurrent network — the reservoir — to perform information-processing tasks. Only the readout weights from the reservoir to the output are trained, leaving the recurrent connections unchanged. This separation makes training computationally efficient while harnessing the expressive power of complex internal dynamics. Echo state networks (ESNs) are the most common implementation.

\item[Second-order phase transition] A phase transition where the system’s behavior changes gradually (continuously) but becomes increasingly sensitive near a critical point. As the control parameter increases, a new pattern of organization (like coordination or synchrony) emerges smoothly. There’s no sudden switch, but near the transition, small fluctuations can have large effects, and the system shows long-range dependencies and slow dynamics.

\item[Self-organized criticality] The tendency of some complex systems to naturally evolve toward a critical state, without the need for fine-tuning external parameters. In this state, the system hovers near a critical point, where activity is balanced between order and randomness, and small events can trigger cascades of varying sizes. In neural systems, this idea is used to explain how spontaneous activity can remain both stable and flexible, supporting efficient information processing.

\item[Stochastic Differential Equations (SDEs)] Differential equations that incorporate random noise terms, extending ordinary differential equations (ODEs) to capture inherent variability in dynamical systems. In computational neuroscience, SDEs model fluctuating neural activity arising from synaptic noise and finite-size effects in neural populations. fWBMs may be formulated as SDEs to capture the stochasticity evident in empirical neural recordings while remaining amenable to gradient-based parameter optimization.

\item[Structural Grounding] A core criterion for fWBMs requiring that model architectures be derived from empirical structural connectivity and regional biological properties, such as macro-scale connectomes and region-specific cell densities.

\item[Thalamo-cortical Motif] A specific neuroanatomical circuit arrangement involving the thalamus and cortex, hypothesized to implement fundamental algorithmic operations like gating or multihead self-attention.

\item[Top-Down Modeling] An approach in computational neuroscience that starts with a high-level cognitive function (e.g., object recognition) and designs a system that can perform that function, often with less emphasis on direct biological implementation at the outset.

\item[Variational Free Energy] A quantity that measures the difference between the brain’s internal model of the world and incoming sensory data. It serves as a tractable proxy for surprise or prediction error, and is minimized during perception and learning under the free energy principle. In this framework, the brain uses variational inference to approximate Bayesian beliefs, with variational free energy guiding the update of those beliefs toward better explanations of sensory input.

\item[Variational Inference] A method for approximating complex probability distributions through optimization. It transforms inference into a tractable learning problem by choosing a simpler family of distributions and minimizing the difference (typically via Kullback-Leibler divergence) between the approximation and the true posterior. In neuroscience, it is hypothesized to provide a tractable mechanism by which the brain can approximate posterior beliefs about the world from noisy sensory input.

\item[Whole-Brain Modeling ] A bottom-up research paradigm that simulates global brain dynamics by connecting multiple neural mass models (representing brain regions) according to an empirical structural connectome.

\end{description}
\end{perspectivebox}
\twocolumn

\printbibliography

@article{myrov2026hierarchical,
  title={Hierarchical whole-brain modeling of critical synchronization dynamics in the human brain},
  author={Myrov, Vladislav and Suleimanova, Alina and Knapi{\v{c}}, Samanta and Partanen, Paula and Vesterinen, Maria and Liu, Wenya and Palva, Satu and Palva, J Matias},
  journal={Proceedings of the National Academy of Sciences},
  volume={123},
  number={12},
  pages={e2505768123},
  year={2026},
  publisher={National Academy of Sciences}
}

@article{fousek2024symmetry,
  title={Symmetry breaking organizes the brain’s resting state manifold},
  author={Fousek, Jan and Rabuffo, Giovanni and Gudibanda, Kashyap and Sheheitli, Hiba and Petkoski, Spase and Jirsa, Viktor},
  journal={Scientific reports},
  volume={14},
  number={1},
  pages={31970},
  year={2024},
  publisher={Nature Publishing Group UK London}
}

@article{gaglioti2026slow,
  title={Slow waves generation and propagation in a model of brain lesions},
  author={Gaglioti, Gianluca and Dalla Porta, Leonardo and Colombo, Michele Angelo and Russo, Simone and Nieus, Thierry and Deco, Gustavo and Corbetta, Maurizio and Sarasso, Simone and Sanchez-Vives, Maria V and Massimini, Marcello},
  journal={NeuroImage},
  pages={121817},
  year={2026},
  publisher={Elsevier}
}

@article{bazinet2023towards,
  title={Towards a biologically annotated brain connectome},
  author={Bazinet, Vincent and Hansen, Justine Y and Misic, Bratislav},
  journal={Nature reviews neuroscience},
  volume={24},
  number={12},
  pages={747--760},
  year={2023},
  publisher={Nature Publishing Group UK London}
}

@article{dalla2025cholinergic,
  title={Cholinergic heterogeneity facilitates synchronization and information flow in a whole-brain model},
  author={Dalla Porta, Leonardo and Fousek, Jan and Destexhe, Alain and Sanchez-Vives, Maria V},
  journal={bioRxiv},
  pages={2025--10},
  year={2025},
  publisher={Cold Spring Harbor Laboratory}
}

@article{sacha2025computational,
  title={A computational approach to evaluate how molecular mechanisms impact large-scale brain activity},
  author={Sacha, Maria and Tesler, Federico and Cofre, Rodrigo and Destexhe, Alain},
  journal={Nature Computational Science},
  volume={5},
  number={5},
  pages={405--417},
  year={2025},
  publisher={Nature Publishing Group US New York}
}

@article{granier2025multihead,
  title={Multihead self-attention in cortico-thalamic circuits},
  author={Granier, Arno and Senn, Walter},
  journal={arXiv preprint arXiv:2504.06354},
  year={2025}
}

@article{safavi2024signatures,
  title={Signatures of criticality in efficient coding networks},
  author={Safavi, Shervin and Chalk, Matthew and Logothetis, Nikos K and Levina, Anna},
  journal={Proceedings of the National Academy of Sciences},
  volume={121},
  number={41},
  pages={e2302730121},
  year={2024},
  publisher={National Academy of Sciences}
}

@article{cambrainha2025criticality,
  title={Criticality at work: scaling in the mouse cortex enhances performance},
  author={Cambrainha, Gustavo G and Castro, Daniel M and de Vasconcelos, Nivaldo AP and Carelli, Pedro V and Copelli, Mauro},
  journal={PRX Life},
  volume={3},
  number={3},
  pages={033026},
  year={2025},
  publisher={APS}
}

@article{haimovici2013brain,
  title={Brain Organization into Resting State Networks Emerges at Criticality on a Model of the Human Connectome},
  author={Haimovici, Ariel and Tagliazucchi, Enzo and Balenzuela, Pablo and Chialvo, Dante R},
  journal={Physical review letters},
  volume={110},
  number={17},
  pages={178101},
  year={2013},
  publisher={APS}
}

@article{rabuffo2025connectome,
  title={The connectome modulates critical brain dynamics across local and global scales},
  author={Rabuffo, Giovanni and Bozzo, Pietro and Nguyen, Bach and Depannemaecker, Damien and Pompili, Marco N and Gollo, Leonardo L and Fukai, Tomoki and Sorrentino, Pierpaolo and Dalla Porta, Leonardo},
  journal={BioRxiv},
  pages={2025--12},
  year={2025},
  publisher={Cold Spring Harbor Laboratory}
}

@article{Hashemi2025,
  author={Hashemi, Meysam and Depannemaecker, Damien and Saggio, Marisa and Triebkorn, Paul and Rabuffo, Giovanni and Fousek, Jan and Ziaeemehr, Abolfazl and Sip, Viktor and Athanasiadis, Anastasios and Breyton, Martin and Woodman, Marmaduke and Wang, Huifang and Petkoski, Spase and Sorrentino, Pierpaolo and Jirsa, Viktor},
  journal={IEEE Reviews in Biomedical Engineering}, 
  title={Principles and Operation of Virtual Brain Twins}, 
  year={2025},
  volume={},
  number={},
  pages={1-29},
  doi={10.1109/RBME.2025.3562951}}

@article{wang2024virtual,
  title={Virtual brain twins: from basic neuroscience to clinical use},
  author={Wang, Huifang E and Triebkorn, Paul and Breyton, Martin and Dollomaja, Borana and Lemarechal, Jean-Didier and Petkoski, Spase and Sorrentino, Pierpaolo and Depannemaecker, Damien and Hashemi, Meysam and Jirsa, Viktor K},
  journal={National Science Review},
  volume={11},
  number={5},
  pages={nwae079},
  year={2024},
  publisher={Oxford University Press}
}

@article{perl2023model,
  title={Model-based whole-brain perturbational landscape of neurodegenerative diseases},
  author={Perl, Yonatan Sanz and Fittipaldi, Sol and Campo, Cecilia Gonzalez and Moguilner, Sebasti{\'a}n and Cruzat, Josephine and Fraile-Vazquez, Matias E and Herzog, Rub{\'e}n and Kringelbach, Morten L and Deco, Gustavo and Prado, Pavel and others},
  journal={Elife},
  volume={12},
  pages={e83970},
  year={2023},
  publisher={eLife Sciences Publications Limited}
}

@article{muller2025critical,
  title={Critical dynamics predicts cognitive performance and provides a common framework for heterogeneous mechanisms impacting cognition},
  author={M{\"u}ller, Paul Manuel and Miron, Gadi and Holtkamp, Martin and Meisel, Christian},
  journal={Proceedings of the National Academy of Sciences},
  volume={122},
  number={14},
  pages={e2417117122},
  year={2025},
  publisher={National Academy of Sciences}
}

@article{bastos2012canonical,
  title={Canonical microcircuits for predictive coding},
  author={Bastos, Andre M and Usrey, W Martin and Adams, Rick A and Mangun, George R and Fries, Pascal and Friston, Karl J},
  journal={Neuron},
  volume={76},
  number={4},
  pages={695--711},
  year={2012},
  publisher={Elsevier}
}

@article{buzsaki2023brain,
  title={Brain rhythms have come of age},
  author={Buzs{\'a}ki, Gy{\"o}rgy and V{\"o}r{\"o}slakos, Mih{\'a}ly},
  journal={Neuron},
  volume={111},
  number={7},
  pages={922--926},
  year={2023},
  publisher={Elsevier}
}

@article{goldman2023simulation,
  title={A comprehensive neural simulation of slow-wave sleep and highly responsive wakefulness dynamics},
  author={Goldman, Jennifer S and Kusch, Lionel and Aquilue, David and Yal{\c{c}}{\i}nkaya, Bahar Hazal and Depannemaecker, Damien and Ancourt, Kevin and Nghiem, Trang-Anh E and Jirsa, Viktor and Destexhe, Alain},
  journal={Frontiers in Computational Neuroscience},
  volume={16},
  pages={1058957},
  year={2023},
  publisher={Frontiers Media SA}
}

@article{nakagawa2014delays,
  title={How delays matter in an oscillatory whole-brain spiking-neuron network model for MEG alpha-rhythms at rest},
  author={Nakagawa, Tristan T and Woolrich, Mark and Luckhoo, Henry and Joensson, Morten and Mohseni, Hamid and Kringelbach, Morten L and Jirsa, Viktor and Deco, Gustavo},
  journal={NeuroImage},
  volume={87},
  pages={383--394},
  year={2014},
  publisher={Elsevier}
}

@article{neftci2019surrogate,
  title={Surrogate gradient learning in spiking neural networks: Bringing the power of gradient-based optimization to spiking neural networks},
  author={Neftci, Emre O and Mostafa, Hesham and Zenke, Friedemann},
  journal={IEEE Signal Processing Magazine},
  volume={36},
  number={6},
  pages={51--63},
  year={2019},
  publisher={IEEE}
}

@article{Mejias2022Mechanisms,
  title={Mechanisms of distributed working memory in a large-scale network of macaque neocortex},
  author={Mej{\'i}as, Jorge F and Wang, Xiao-Jing},
  journal={elife},
  volume={11},
  pages={e72136},
  year={2022},
  publisher={eLife Sciences Publications Limited}
}

@article{Goulas2021Bio,
  title={Bio-instantiated recurrent neural networks: Integrating neurobiology-based network topology in artificial networks},
  author={Goulas, Alexandros and Damicelli, Fabrizio and Hilgetag, Claus C},
  journal={Neural Networks},
  volume={142},
  pages={608--618},
  year={2021},
  publisher={Elsevier}
}

@article{Damicelli2022Brain,
  title={Brain connectivity meets reservoir computing},
  author={Damicelli, Fabrizio and Hilgetag, Claus C and Goulas, Alexandros},
  journal={PLoS Computational Biology},
  volume={18},
  number={11},
  pages={e1010639},
  year={2022},
  publisher={Public Library of Science San Francisco, CA USA}
}

@article{OByrne2022Critical,
  title={How critical is brain criticality?},
  author={O’Byrne, Jordan and Jerbi, Karim},
  journal={Trends in Neurosciences},
  volume={45},
  number={11},
  pages={820--837},
  year={2022},
  publisher={Elsevier}
}

@article{Chialvo2010Emergent,
  title={Emergent complex neural dynamics},
  author={Chialvo, Dante R},
  journal={Nature physics},
  volume={6},
  number={10},
  pages={744--750},
  year={2010},
  publisher={Nature Publishing Group UK London}
}

@article{Williams2021Epistemic,
  title={Epistemic irrationality in the Bayesian brain},
  author={Williams, Daniel},
  journal={The British Journal for the Philosophy of Science},
  year={2021},
  publisher={The University of Chicago Press}
}

@article{Sprevak2024Predictive,
  title={Predictive coding I: Introduction},
  author={Sprevak, Mark},
  journal={Philosophy Compass},
  volume={19},
  number={1},
  pages={e12950},
  year={2024},
  publisher={Wiley Online Library}
}

@article{Effenberger2025Functional,
  title={The functional role of oscillatory dynamics in neocortical circuits: a computational perspective},
  author={Effenberger, Felix and Carvalho, Pedro and Dubinin, Igor and Singer, Wolf},
  journal={Proceedings of the National Academy of Sciences},
  volume={122},
  number={4},
  pages={e2412830122},
  year={2025},
  publisher={National Academy of Sciences}
}

@article{Grytskyy2021Learning,
  title={A learning rule balancing energy consumption and information maximization in a feed-forward neuronal network},
  author={Grytskyy, Dmytro and Jolivet, Renaud B},
  journal={arXiv preprint arXiv:2103.06562},
  year={2021}
}

@article{KriegeskorteDouglas2018,
  author  = {Kriegeskorte, Nikolaus and Douglas, Rodney},
  title   = {Cognitive computational neuroscience},
  journal = {Nature Neuroscience},
  year    = {2018},
  volume  = {21},
  number  = {9},
  pages   = {1148--1160},
  doi     = {10.1038/s41593-018-0210-5}
}

@article{Bellec2020Solution,
    author    = {Bellec, Guillaume and Scherr, Franz and Subramoney, Anand and Hajek, Elias and Salaj, Darjan and Legenstein, Robert and Maass, Wolfgang},
    title     = {A solution to the learning dilemma for recurrent networks of spiking neurons},
  journal   = {Nature Communications},
    year      = {2020},
    volume    = {11},
    number    = {1},
    pages     = {3625}
}

@article{Bjaalie2008understanding,
    author    = {Bjaalie, Jan},
    title     = {Understanding the brain through neuroinformatics},
    journal   = {Frontiers in Neuroscience},
    year      = {2008},
    volume    = {2},
    doi       = {10.3389/neuro.01.022.2008}
}

@article{Bottemanne2025BayesianBrain,
    author    = {Bottemanne, Hugo},
    title     = {Bayesian brain theory: Computational neuroscience of belief},
    journal   = {Neuroscience},
    year      = {2025},
    volume    = {566},
    pages     = {198--204},
    doi       = {10.1016/j.neuroscience.2024.12.003}
}

@article{Goldstein2025UnifiedAcoustic,
  title={A unified acoustic-to-speech-to-language embedding space captures the neural basis of natural language processing in everyday conversations},
  author={Goldstein, Ariel and Wang, Haocheng and Niekerken, Leonard and Schain, Mariano and Zada, Zaid and Aubrey, Bobbi and Sheffer, Tom and Nastase, Samuel A and Gazula, Harshvardhan and Singh, Aditi and others},
  journal={Nature human behaviour},
  pages={1--15},
  year={2025},
  publisher={Nature Publishing Group UK London}
}

@article{VanEssen2012Human,
  title={The Human Connectome Project: a data acquisition perspective},
  author={Van Essen, David C and Ugurbil, Kamil and Auerbach, Edward and Barch, Deanna and Behrens, Timothy EJ and Bucholz, Richard and Chang, Acer and Chen, Liyong and Corbetta, Maurizio and Curtiss, Sandra W and others},
  journal={Neuroimage},
  volume={62},
  number={4},
  pages={2222--2231},
  year={2012},
  publisher={Elsevier}
}

@article{Bowers2022DeepProblemsNNVision,
    author    = {Bowers, Jeffrey S. and Malhotra, Gaurav and Dujmovi{\'c}, Marko and Montero, M{\'o}nica L. and Tsvetkov, Christo and Biscione, Valerio and Puebla, Guillermo and Adolfi, Federico and Hummel, John E. and Heaton, Ryan F. and Evans, Ben D. and Mitchell, Rebecca and Blything, Ryan},
    title     = {Deep problems with neural network models of human vision},
    journal   = {Behavioral and Brain Sciences},
    year      = {2022},
    volume    = {46},
    pages     = {e385},
    doi       = {10.1017/S0140525X22002813}
}

@article{Breakspear2017DynamicModels,
    author    = {Breakspear, Michael},
    title     = {Dynamic models of large-scale brain activity},
    journal   = {Nature Neuroscience},
    year      = {2017},
    volume    = {20},
    number    = {3},
    pages     = {340--352},
    doi       = {10.1038/nn.4497}
}

@article{DeLeeuwDeep,
   title={Deep biophysical modeling: Using cortical columns as neural network nodes},
   author={de Leeuw, Dasja and Goebel, Rainer and Senden, Mario},
   note={in preparation}
}

@article{WeidlerLateral,
   title={Lateral Prefrontal Cortex Builds and Distributes Action Plans},
   author={Weidler, Tonio and Goebel, Rainer and Senden, Mario},
   note={in preparation}
}

@article{Chen2018NeuralODE,
    author    = {Chen, Ricky T. Q. and Rubanova, Yulia and Bettencourt, Jesse and Duvenaud, David},
    title     = {Neural Ordinary Differential Equations},
    journal   = {arXiv preprint arXiv:1806.07366},
    year      = {2018},
    doi       = {10.48550/arXiv.1806.07366}
}

@article{Chirimuuta2013MinimalModels,
    author    = {Chirimuuta, Mazviita},
    title     = {Minimal models and canonical neural computations: the distinctness of computational explanation in neuroscience},
    journal   = {Synthese},
    year      = {2013},
    volume    = {191},
    number    = {2},
    pages     = {127--153},
    doi       = {10.1007/s11229-013-0369-y}
}

@article{Rao1999Predictive,
  title={Predictive coding in the visual cortex: a functional interpretation of some extra-classical receptive-field effects},
  author={Rao, Rajesh PN and Ballard, Dana H},
  journal={Nature neuroscience},
  volume={2},
  number={1},
  pages={79--87},
  year={1999},
  publisher={Nature Publishing Group}
}

@article{Friston2010FreeEnergy,
  title={The free-energy principle: a unified brain theory?},
  author={Friston, Karl},
  journal={Nature reviews neuroscience},
  volume={11},
  number={2},
  pages={127--138},
  year={2010},
  publisher={Nature publishing group}
}

@book{Beggs2022CriticalCortex,
  title={The cortex and the critical point: understanding the power of emergence},
  author={Beggs, John M},
  year={2022},
  publisher={MIT Press}
}

@article{Deco2012Anatomy,
  title={How anatomy shapes dynamics: a semi-analytical study of the brain at rest by a simple spin model},
  author={Deco, Gustavo and Senden, Mario and Jirsa, Viktor},
  journal={Frontiers in computational neuroscience},
  volume={6},
  pages={68},
  year={2012},
  publisher={Frontiers Media SA}
}

@article{Senden2014RichClub,
  title={Rich club organization supports a diverse set of functional network configurations},
  author={Senden, Mario and Deco, Gustavo and De Reus, Marcel A and Goebel, Rainer and Van Den Heuvel, Martijn P},
  journal={Neuroimage},
  volume={96},
  pages={174--182},
  year={2014},
  publisher={Elsevier}
}

@article{Deco2011Emerging,
    author    = {Deco, Gustavo and Jirsa, Viktor K. and McIntosh, Anthony R.},
    title     = {Emerging concepts for the dynamical organization of resting-state activity in the brain},
    journal   = {Nature Reviews Neuroscience},
    year      = {2011},
    volume    = {12},
    number    = {1},
    pages     = {43--56},
    doi       = {10.1038/nrn2961}
}

@article{Deco2012RestingCriticality,
    author    = {Deco, Gustavo and Jirsa, Viktor K.},
    title     = {Ongoing cortical activity at rest: criticality, multistability, and ghost attractors},
    journal   = {Journal of Neuroscience},
    year      = {2012},
    volume    = {32},
    number    = {10},
    pages     = {3366--3375},
    doi       = {10.1523/JNEUROSCI.2523-11.2012}
}

@article{DecoKringelbach2014Great,
    author    = {Deco, Gustavo and Kringelbach, Morten L.},
    title     = {Great expectations: Using whole-brain computational connectomics for understanding neuropsychiatric disorders},
    journal   = {Neuron},
    year      = {2014},
    volume    = {84},
    number    = {5},
    pages     = {892--905},
    doi       = {10.1016/j.neuron.2014.08.034}
}

@article{Djurfeldt2008LargeScaleModeling,
    author    = {Djurfeldt, Mikael},
    title     = {Large-scale modeling – a tool for conquering the complexity of the brain},
    journal   = {Frontiers in Neuroinformatics},
    year      = {2008},
    volume    = {2},
    doi       = {10.3389/neuro.11.001.2008}
}

@article{Doerig2023Neuroconnectionist,
    author    = {Doerig, Adrien and Sommers, Rowan P and Seeliger, Katja and Richards, Blake and Ismael, Jenann and Lindsay, Grace W and Kording, Konrad P and Konkle, Talia and Van Gerven, Marcel AJ and Kriegeskorte, Nikolaus and others},
    title     = {The neuroconnectionist research programme},
    journal   = {Nature Reviews Neuroscience},
    year      = {2023},
    volume    = {24},
    number    = {7},
    pages     = {431--450}
}

@article{Eliasmith2014LargeScaleModels,
    author    = {Eliasmith, Chris and Trujillo, Oliver},
    title     = {The use and abuse of large-scale brain models},
    journal   = {Current Opinion in Neurobiology},
    year      = {2014},
    volume    = {25},
    pages     = {1--6},
    doi       = {https://doi.org/10.1016/j.conb.2013.09.009}
}

@article{Friston2005CorticalResponses,
    author    = {Friston, Karl},
    title     = {A theory of cortical responses},
    journal   = {Philosophical Transactions of the Royal Society B: Biological Sciences},
    year      = {2005},
    volume    = {360},
    number    = {1456},
    pages     = {815--836},
    doi       = {10.1098/rstb.2005.1622}
}

@article{Friston2009FreeEnergy,
    author    = {Friston, Karl and Kiebel, Stefan},
    title     = {Predictive coding under the free-energy principle},
    journal   = {Philosophical Transactions of the Royal Society B: Biological Sciences},
    year      = {2009},
    volume    = {364},
    number    = {1521},
    pages     = {1211--1221},
    doi       = {10.1098/rstb.2008.0300}
}

@inproceedings{Grauman2022Ego4D,
    author    = {Grauman, Kristen and Westbury, Andrew and Byrne, Eugene and Chavis, Zachary and Furnari, Antonino and Girdhar, Rohit and Hamburger, Jackson and Jiang, Hao and Liu, Miao and Liu, Xingyu and others},
    title     = {Ego4d: Around the world in 3,000 hours of egocentric video},
    booktitle = {Proceedings of the IEEE/CVF conference on computer vision and pattern recognition},
    year      = {2022},
    pages     = {18995--19012}
}

@incollection{Griffiths2021WholeBrain,
    author    = {Griffiths, John D. and Bastiaens, Sorenza P. and Kaboodvand, Neda},
    title     = {Whole-brain modelling: Past, present, and future},
    booktitle = {Computational Modelling of the Brain},
    year      = {2021},
    pages     = {313--355},
    doi       = {10.1007/978-3-030-89439-9_13}
}

@article{Hengen2024Criticality,
    author    = {Hengen, Keith B. and Shew, Woodrow L.},
    title     = {Is criticality a unified set-point of brain function?},
    journal   = {bioRxiv},
    year      = {2024},
    doi       = {10.1101/2024.09.02.610815}
}

@article{Hesse2014Criticality,
    author    = {Hesse, Janina and Gross, Thilo},
    title     = {Self-organized criticality as a fundamental property of neural systems},
    journal   = {Frontiers in Systems Neuroscience},
    year      = {2014},
    volume    = {8},
    doi       = {10.3389/fnsys.2014.00166}
}

@article{Honey2009Predicting,
    author    = {Honey, C. J. and Sporns, O. and Cammoun, L. and Gigandet, X. and Thiran, J. P. and Meuli, R. and Hagmann, P.},
    title     = {Predicting human resting-state functional connectivity from structural connectivity},
    journal   = {Proceedings of the National Academy of Sciences of the United States of America},
    year      = {2009},
    volume    = {106},
    number    = {6},
    pages     = {2035--2040},
    doi       = {10.1073/pnas.0811168106}
}

@article{Hosseini2020HierarchicalPredictiveCoding,
    author    = {Hosseini, Matin and Maida, Anthony S.},
    title     = {Hierarchical predictive coding models in a deep-learning framework},
    journal   = {arXiv preprint arXiv:2005.03230},
    year      = {2020}
}

@article{Kubilius2018CORnet,
    author    = {Kubilius, Jonas and Schrimpf, Martin and Nayebi, Aran and Bear, Daniel and Yamins, Daniel L. K. and DiCarlo, James J.},
    title     = {CORnet: Modeling the Neural Mechanisms of Core Object Recognition},
    journal   = {bioRxiv},
    year      = {2018},
    doi       = {10.1101/408385}
}

@article{Lange2023BayesianCoding,
    author    = {Lange, Richard D. and Shivkumar, Sabyasachi and Chattoraj, Ankani and Haefner, Ralf M.},
    title     = {Bayesian encoding and decoding as distinct perspectives on neural coding},
    journal   = {Nature Neuroscience},
    year      = {2023},
    volume    = {26},
    number    = {12},
    pages     = {2063--2072},
    doi       = {10.1038/s41593-023-01458-6}
}

@article{Liu2019NeuralSDE,
    author    = {Liu, Xuanqing and Xiao, Tesi and Si, Si and Cao, Qin and Kumar, Sanjiv and Hsieh, Cho-Jui},
    title     = {Neural SDE: Stabilizing Neural ODE Networks with Stochastic Noise},
    journal   = {arXiv preprint arXiv:1906.02355},
    year      = {2019},
    doi       = {10.48550/arXiv.1906.02355}
}

@article{Luo2024multimodal,
    author    = {Luo, Na and Shi, Weiyang and Yang, Zhengyi and Song, Ming and Jiang, Tianzi},
    title     = {Multimodal Fusion of Brain Imaging Data: Methods and Applications},
    journal   = {Machine Intelligence Research},
    year      = {2024},
    volume    = {21},
    number    = {1},
    pages     = {136--152},
    doi       = {10.1007/s11633-023-1442-8}
}

@article{Machens2012Building,
    author    = {Machens, Christian K},
    title     = {Building the human brain},
    journal   = {Science},
    year      = {2012},
    volume    = {338},
    number    = {6111},
    pages     = {1156}
}

@article{Piccinini2014Foundations,
    author    = {Piccinini, Gualtiero and Shagrir, Oron},
    title     = {Foundations of computational neuroscience},
    journal   = {Current Opinion in Neurobiology},
    year      = {2014},
    volume    = {25},
    pages     = {25--30},
    doi       = {10.1016/j.conb.2013.10.005}
}

@article{Pronold2024MultiscaleModel,
    author    = {Pronold, Jari and van Meegen, Alexander and Shimoura, Renan O. and Vollenbröker, Hannah and Senden, Mario and Hilgetag, Claus C. and Bakker, Rembrandt and van Albada, Sacha J.},
    title     = {Multi-scale spiking network model of human cerebral cortex},
    journal   = {Cerebral Cortex},
    year      = {2024},
    volume    = {34},
    number    = {10},
    doi       = {10.1093/cercor/bhae409}
}

@article{Richards2019Deep,
    author    = {Richards, Blake A and Lillicrap, Timothy P and Beaudoin, Philippe and Bengio, Yoshua and Bogacz, Rafal and Christensen, Amelia and Clopath, Claudia and Costa, Rui Ponte and de Berker, Archy and Ganguli, Surya and others},
    title     = {A deep learning framework for neuroscience},
    journal   = {Nature neuroscience},
    year      = {2019},
    volume    = {22},
    number    = {11},
    pages     = {1761--1770}
}

@article{Rood2025HumanCellAtlas,
    author    = {Rood, Jennifer E. and Wynne, Samantha and Robson, Lucia and Hupalowska, Anna and Randell, John and Teichmann, Sarah A. and Regev, Aviv},
    title     = {The Human Cell Atlas from a cell census to a unified foundation model},
    journal   = {Nature},
    year      = {2025},
    volume    = {637},
    pages     = {1065--1071}
}

@article{Salvatori2023BrainInspired,
    author    = {Salvatori, Tommaso and Mali, Ankur and Buckley, Christopher L. and Lukasiewicz, Thomas and Rao, Rajesh P.N. and Friston, Karl and Ororbia, Alexander},
    title     = {Brain-inspired computational intelligence via predictive coding},
    journal   = {arXiv preprint arXiv:2308.07870},
    year      = {2023}
}

@article{Senden2023ModularIntegrative,
    author    = {Senden, Mario and van Albada, Sacha J and Pezzulo, Giovanni and Falotico, Egidio and Hashim, Ibrahim and Kroner, Alexander and Kurth, Anno C and Lanillos, Pablo and Narayanan, Vaishnavi and Pennartz, Cyriel and Petrovici, Mihai A and Steffen, Lea and Weidler, Tonio and Goebel, Rainer},
    title     = {Modular-integrative modeling: a new framework for building brain models that blend biological realism and functional performance},
    journal   = {National Science Review},
    year      = {2023},
    volume    = {11},
    number    = {5},
    pages     = {nwad318},
    doi       = {10.1093/nsr/nwad318}
}

@article{Senden2026EvolvingLandscape,
  title={The evolving landscape of neuroscience},
  author={Senden, Mario},
  journal={Aperture Neuro},
  volume={6},
  number={SI 2},
  pages={156380},
  year={2026},
  doi={10.52294/001c.156380}
}

@article{Vohryzek2022BrainStates,
    author    = {Vohryzek, Jakub and Cabral, Joana and Vuust, Peter and Deco, Gustavo and Kringelbach, Morten L.},
    title     = {Understanding brain states across spacetime informed by whole-brain modelling},
    journal   = {Philosophical Transactions of the Royal Society A},
    year      = {2022},
    volume    = {380},
    number    = {2227},
    doi       = {10.1098/rsta.2021.0247}
}

@article{Wunderlich2021EventBased,
    author    = {Wunderlich, T. C. and Pehle, C.},
    title     = {Event-based backpropagation can compute exact gradients for spiking neural networks},
    journal   = {Scientific Reports},
    year      = {2021},
    volume    = {11},
    number    = {1},
    pages     = {12156}
}

@article{Yamins2016GoalDriven,
    author    = {Yamins, Daniel L. K. and DiCarlo, James J.},
    title     = {Using goal-driven deep learning models to understand sensory cortex},
    journal   = {Nature Neuroscience},
    year      = {2016},
    volume    = {19},
    number    = {3},
    pages     = {356--365},
    doi       = {10.1038/nn.4244}
}

@article{Zhou2006Hierarchical,
  author  = {Zhou, C. S. and Zemanov\'a, L. and Zamora, G. and Hilgetag, C. C. and Kurths, J.},
  title   = {Hierarchical Organization Unveiled by Functional Connectivity in Complex Brain Networks},
  journal = {Physical Review E},
  year    = {2006},
  volume    = {97},
  number    = {},
  pages     = {238103},
  doi       = {10.1103/PhysRevLett.97.238103}
}

@article{Gomez2010Synchronization,
  author  = {G\'omez-Garde\~{n}es, J. and Zamora-L\'opez, G. and Moreno, Y. and Arenas, A.},
  title   = {From Modular to Centralized Organization of Synchronization in Functional Areas of the Cat Cerebral Cortex},
  journal = {PLoS ONE},
  year    = {2010},
  volume    = {5},
  number    = {8},
  pages     = {e12313},
  doi       = {10.1371/journal.pone.0012313}
}

@article{Perez2020Sleep,
  author  = {Perez Ipi\~{n}a, I. and Donnelly Kehoe, P. and Kringelbach, M. and Laufs, H. and Iba\~{n}ez, A. and Deco, G. and Sanz Perl, Y. and Tagliazucchi, E.},
  title   = {Modeling regional changes in dynamic stability during sleep and wakefulness},
  journal = {NeuroImage},
  year    = {2020},
  volume    = {215},
  number    = {},
  pages     = {116833},
  doi       = {10.1016/j.neuroimage.2020.116833}
}

@article{Lopez2021Stability,
  author  = {L\'opez-Gonz\'alez, A. and Panda, R. and Ponce-Alvarez, A. and Zamora-L\'opez, G. and Escrichs, A. and Martial, C. and Thibaut, A. and Gosseries, O. and Kringelbach, M. and Annen, A. and Laureys, S. and Deco, G.},
  title   = {Loss of consciousness reduces the stability of brain hubs and the heterogeneity of brain dynamics},
  journal = {Communications Biology},
  year    = {2021},
  volume    = {4},
  number    = {},
  pages     = {1037},
  doi       = {10.1038/s42003-021-02537-9 | }
}

@article{Patow2023Alzheimers,
  author  = {Patow, G. and Stefanovski, L. and Ritter, P. and Deco, G. and Kobeleva, X.},
  title   = {Whole‑brain modeling of the differential influences of amyloid‑beta and tau in Alzheimer’s disease},
  journal = {Alzheimer's Research \& Therapy},
  year    = {2023},
  volume    = {15},
  number    = {},
  pages     = {210},
  doi       = {10.1186/s13195-023-01349-9}
}

@article{Dallmer2023Epilepsy,
  author  = {Dallmer-Zerbe, I. and Jiruska, P. and Hlinka, J.},
  title   = {Personalized dynamic network models of the human brain as a future tool for planning and optimizing epilepsy therapy},
  journal = {Epilepsia},
  year    = {2023},
  volume    = {64},
  number    = {},
  pages     = {2221-2238},
  doi       = {10.1111/epi.17690}
}

@article{Hagmann2008Structural,
  author  = {Hagmann, Patric and Cammoun, Leila and Gigandet, Xavier and Meuli, Reto and Honey, Christopher J. and Wedeen, Van J. and Sporns, Olaf},
  title   = {Mapping the structural core of human cerebral cortex},
  journal = {PLoS Biology},
  year    = {2008},
  volume  = {6},
  number  = {7},
  pages   = {e159},
  doi     = {10.1371/journal.pbio.0060159}
}

@article{Rubinov2010Complex,
  author  = {Rubinov, Mikail and Sporns, Olaf},
  title   = {Complex network measures of brain connectivity: Uses and interpretations},
  journal = {NeuroImage},
  year    = {2010},
  volume  = {52},
  number  = {3},
  pages   = {1059--1069},
  doi     = {10.1016/j.neuroimage.2009.10.003}
}

@article{Deneve2016EfficientCodes,
  author  = {Den{\`e}ve, Sophie and Machens, Christian K.},
  title   = {Efficient codes and balanced networks},
  journal = {Nature Neuroscience},
  year    = {2016},
  volume  = {19},
  number  = {3},
  pages   = {375--382},
  doi     = {10.1038/nn.4243}
}

@article{VanHolk2024CogFlex,
  author  = {van Holk, Maya and Mej{\'i}as, Jorge F.},
  title   = {Biologically plausible models of cognitive flexibility: merging recurrent neural networks with full-brain dynamics},
  journal = {Current Opinion in Behavioral Sciences},
  year    = {2024},
  volume  = {56},
  pages   = {101351},
  doi     = {10.1016/j.cobeha.2024.101351}
}

@article{Palmigiano2017Routing,
  author  = {Palmigiano, Agostina and Geisel, Theo and Wolf, Fred and Battaglia, Demian},
  title   = {Flexible information routing by transient synchrony},
  journal = {Nature Neuroscience},
  year    = {2017},
  volume  = {20},
  number  = {7},
  pages   = {1014--1022},
  doi     = {10.1038/nn.4569}
}
\end{document}